\documentclass[12pt]{article}
\usepackage{amsmath}
\usepackage{graphicx}
\usepackage{enumerate}
\usepackage{natbib}
\usepackage{url} 
\usepackage{amssymb}
\usepackage{bbm}
\usepackage{psfrag,epsf}
\usepackage{natbib}
\usepackage{tabularx}
\usepackage{url} 
\usepackage{tikz}
\usepackage{blkarray}
\usetikzlibrary{calc,arrows,positioning,decorations.pathreplacing}
\usepackage{bm}
\usepackage{float}
\usepackage{multirow}
\usepackage{threeparttable}
\usepackage{booktabs}
\usepackage{dsfont}
\usepackage{authblk}
\usepackage{algorithm2e}
\RestyleAlgo{ruled}
\usepackage{tikz-network}
\usepackage{amsthm}
\theoremstyle{plain}
\newtheorem{prop}{Proposition}
\newtheorem*{proof*}{Proof}
\newtheorem{assum}{Assumption}
\newtheorem{defi}{Definition}

\newtheorem{theor}{Theorem}
\usepackage{color}

\begin{document}

\def\spacingset#1{\renewcommand{\baselinestretch}%
{#1}\small\normalsize} \spacingset{1}


{
  \title{\bf Estimating hidden population size from a single respondent-driven sampling survey}
  \author[1]{Mamadou Yauck\thanks{
    The authors gratefully acknowledge \textit{the Natural Sciences and Engineering Research
Council of Canada (RGPIN-2022-03309) and the Fonds de Recherche du Québec - Santé, Programme
de bourses de chercheur-boursier Junior 1 - Intelligence artificelle et santé numérique
(322200).}}\hspace{.2cm} }
\author[2]{Erica EM Moodie}
\author[3]{Alain Fourmigue}
\author[3]{Milada Dvorakova}
\author[4]{Gilles Lambert}
\author[5]{Daniel Grace}
\author[2]{Joseph Cox}

\affil[1]{Department of Mathematics, Université du Québec à Montréal, Montréal, Québec, Canada}
\affil[2]{Department of Epidemiology, Biostatistics and Occupational Health, McGill University, Montréal, Quebec, Canada}
\affil[3]{Research Institute of the McGill University Health Centre, Montréal, Québec, Canada}
\affil[4]{Institut national de santé publique du Québec, Montréal, Québec, Canada}
\affil[5]{Dalla Lana School of Public Health, University of Toronto}
  \maketitle
} 


\bigskip
\begin{abstract}
This work is concerned with the estimation of hard-to-reach population sizes using a single respondent-driven sampling (RDS) survey, a variant of chain-referral sampling that leverages social relationships to reach members of a hidden population. The popularity of RDS as a standard approach for surveying hidden populations brings theoretical and methodological challenges regarding the estimation of population sizes, mainly for public health purposes. This paper proposes a frequentist, model-based framework for estimating the size of a hidden population using a network-based approach. An optimization algorithm is proposed for obtaining the identification region of the target parameter when model assumptions are violated. We characterize the asymptotic behavior of our proposed methodology and assess its finite sample performance under departures from model assumptions.
\end{abstract}

\noindent%
{\it Keywords:}  chain-referral sampling; graph theory; identification bounds; respondent-driven sampling; social networks; stochastic optimization.
\vfill

\newpage
\spacingset{1.45} 
\section{Introduction}
\label{sec:intro}
\addtolength{\textheight}{.5in}%
\textit{Respondent-driven sampling} (RDS) is a variant of link-tracing sampling that relies on social interactions to recruit individuals from a hard-to-reach population, for which (i) there is no sampling frame and (ii) membership may be (but is not necessarily) associated with stigmatized behavior \citep{Hec97}. The RDS recruitment process is usually carried out over a number of \textit{waves}, where each recruited individual receives a number of \textit{coupons} and is asked to recruit individuals from the target population with whom they share social relationships, starting from initial \textit{seed} participants; recruiters receive (monetary and/or non-monetary incentives) for each distributed coupon when the selected participant, often termed a \textit{neighbour} in reference to an adjoining node in a social network, agrees to take part in the study. RDS provides many advantages over traditional chain-referral sampling methods. First, the RDS process occurs through many waves, assuring recruitment beyond initial seeds and greater heterogeneity. Allowing participants to partially control the process by inviting their peers helps to reduce privacy concerns for participants who are not willing to share their personal network of contacts. Finally, asking for participants to report on the size of their network within the target population allows for post-recruitment adjustments, as it is generally believed that individuals with larger personal networks are more likely to appear in the sample.

RDS has gained popularity over the last twenty years as a standard sampling tool for surveying hard-to-reach populations (\citealt{platt2006methods,world2010guidelines, world2016consolidated}). The early literature on statistical methodology for analyzing RDS data has mainly focused on the estimation of population means and proportions (\citealt{Vol08,RDSvar-2018}). The growing interest in estimating the size of hard-to-reach populations such as people who inject drugs, sex workers, or gay, bisexual and other men who have sex with men (GBM), driven by the need to provide accurate information regarding the size and the structure of those populations for public health policy purposes, has contributed to important theoretical and methodological advances in the statistical literature.

Current approaches for estimating the size $N$ of a hard-to-reach population either rely on a multiplier method (\citealt{heimer2010estimation, paz2011many}) or approximations to the RDS design. \cite{handcock2015estimating} developed a technique for estimating $N$ by assuming that the RDS process behaves as a successive, without-replacement sampling process where the probability of appearing in the sample is proportional to the size of an individual's personal network, or \textit{degree}. They assumed that the average degree should decrease with increasing recruitment waves, and that the rate of such decrease reveals some information about $N$. Recently, \cite{crawford2018hidden} proposed a network-based approach to estimating population sizes by acknowledging that the RDS recruitment chain provides only a partial picture of the underlying population network, and by assuming that for a given individual, the number of unrecruited peers with whom they share social ties just before their recruitment is independent of the design and follows a distribution that includes $N$ as a parameter. By further leveraging recruitment times and population degrees, \cite{crawford2018hidden} developed a Bayesian framework which provides estimates for the number of connections between recruited and unrecruited individuals. More recently, \cite{kim2021population} considered multiple RDS studies and proposed a population size estimator using capture-recapture methods and by modeling the RDS design as a successive sampling process.

This paper is the first to propose a model-based frequentist approach to the estimation of population sizes using a single RDS survey. First, we propose a network-based approach to characterize the underlying structure of the target population, i.e., its size and network density, with the intuition that given the number of coupons received by a new recruit, the chance of successfully distributing all coupons and receiving the associated incentive should depend on the number of unrecruited neighbours at the time of recruitment, as well as the chance that the recruited individuals take their coupons to the surveyer for participation in the study.

Section \ref{sec:design} introduces some concepts of graph theory for describing the underlying structure of the population. In Section \ref{sec:methodo}, we present a methodology for estimating $N$ under graphical assumptions about the target population and about the decision-making of both recruiters and potential recruits regarding the distribution of coupons. In Section \ref{sec:identbound}, we propose a stochastic optimization algorithm for deriving identification bounds for $N$ when model assumptions are violated. The performance of this new methodology is investigated in Section \ref{sec:simulation}, and then applied to a canonical RDS study frequently cited in the methodological literature as well as a more recent RDS study conducted in Canada in Section \ref{sec:casestudy}.

\section{Sampling design}\label{sec:design}
Let $\mathcal{P}=\{1,\dots, N\}$ be the set of units in the population, where $N$ is the size of the closed population. We assume that the population has an underlying network structure represented by a \textit{graph} $G=(V,E)$, where $V$ is the set of nodes or vertices, with $|V|=N$, and $E$ denotes the set of links or edges connecting nodes. The ordered pair $(u, v)$ represents an edge between nodes $u, v \in V$. When the graph is \textit{undirected}, $(u, v) \in E$ implies $(v, u) \in E$; in the case of \textit{directed} graphs, $(u, v) \in E$ does not imply $(v, u) \in E$. {\it Adjacent} vertices are linked by an edge; a sequence of distinct vertices that are adjacent is called a {\it path}. Two adjacent vertices are said to be \textit{neighbours}. A \textit{connected} graph is a graph for which there is a path from any node to any other node in the network \citep{west2017introduction}. Let $\bm{A}$ be the $N\times N$ adjacency matrix for $G$ with elements $A_{jk}=1$ if $(j, k) \in E$ and $A_{jk}=0$ otherwise, with $A_{jj}=0$. The degree for the $i$th unit $d_{i}=\sum_{k=1}^N A_{ij}$ represents the number of links they share with other units in the population.

A typical link-tracing sampling procedure starts with the selection of initial \textit{seed} participants from the target population; in practice, seeds could be chosen either by convenience or in a way that does not induce correlation with observed and/or unobserved trait values or network characteristics. The initial recruits, now recruiters, are then tasked with recruiting additional participants by reaching out through existing links with unrecruited neighbouring units. This paper explores a sampling design that starts with the selection of initial participants, and which allows each current recruit to sequentially add more units through a recruitment system that specifies the maximum number of participants one is allowed to bring into the study. The sampling process is formally described as follows.
\begin{itemize}
\item Step 1. In the first stage of recruitment, \textit{wave zero}, a fixed number of individuals are selected as seeds from the target population.
\item Step 2. Initial seed participants from Step 1 recruit a fixed number of individuals from among their unrecruited neighbours; for each seed, the number of recruits is bounded above by the number of allocated coupons.
\item Step 3. Recruits from Step 2 are then given coupons with the task of recruiting among their neighbours. Through multiple recruitment waves, the process continues until a sample size $n$ is reached.
\end{itemize}
By using a system of coupons to uniquely identify each new participant's recruiter, one can construct a recruitment subgraph illustrating both the sample propagation through the network and recruiter-recruit relationships.
\begin{defi}{\textbf{(The recruitment subgraph).}}\label{defi:recsubgraph}
The recruitment subgraph is a directed graph $G_T=(V_T, E_T)$, where $V_T\subset V$ is the set of sampled nodes, with $|V_T|=n$, and $E_T\subset E$ is the set of edges, with $\{u, v\} \in E_T$ if $u$ recruited $v$ into the study.
\end{defi}
The recruitment subgraph does not report any links between any two nodes with no shared recruitment history. Next, we define an augmented recruitment subgraph which contains all unobserved links within the sample. Let $V_P=\{i \notin V_T: \exists \,j \in V_T \,\,\text{and}\,\, (i,j) \in E \}$ represent the set of unrecruited units connected to at least one unit in the RDS recruitment graph, and $E_P=\{ (i,j): i \in V_T,\,j \in V_P\,\, \text{and}\,\, (i,j) \in E\}$ is the set of ties connecting units in $V_P$ to units in $V_T$. Next, we define the graph induced by the union of $V_P$ and $V_T$.
\begin{defi}{\textbf{(Augmented recruitment-induced subgraph).}}\label{defi:recaugsubgraph}
The augmented recruitment-induced subgraph is the undirected graph $G_{U}=(V_U,E_U)$ with $V_U=V_T\cup V_P$ and $E_U=E_R\cup E_P$.
\end{defi}

 \begin{figure}[H]
\begin{center}
\includegraphics[scale=.95]{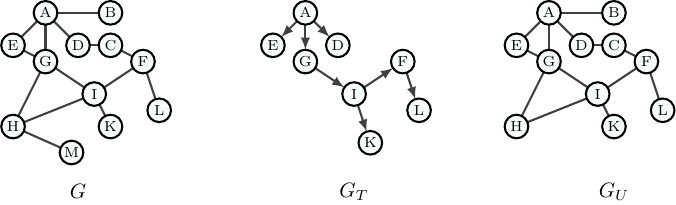}
\caption{Illustration of the population graph $G$, the recruitment subgraph $G_T$ and the augmented recruitment subgraph $G_U$.}
\label{fig:RDSgraphs}
\end{center}
\end{figure}

We assume that recruited units accurately report on their population degrees $d_i$, $i=1, \dots, n$. Let $t_i$ denote the recruitment time for the $i$th individual, $i=1, \dots, n$, where $t_1<t_2<\dots<t_n$. The observed recruitment data are $\bm{D}=\{G_T, d_i, t_i; i=1,\dots, n\}$. The recruitment graphs, along with the population graph, are illustrated in Figure \ref{fig:RDSgraphs}. While $G_U$ cannot be observed in a real-world RDS setting, this subgraph proves useful for describing the topology of the latent portion of $G_T$; note that the proposed methodology does not assume the analyst has access to this subgraph.

\section{Methodology}\label{sec:methodo}
Statistical inference for $N$ in a design-based frequentist setting for RDS assumes knowledge of the parameters that govern the recruitment process through, for example, additional assumptions about recruiters' decision-making. In this paper, we make a set of assumptions that will contribute to uncovering the unobserved fraction of the population graph $G$ indexed by the recruitment subgraph $G_T$. This section introduces assumptions regarding the topology of the population graph and the sampling design, presents a model for the observed data and provides an inferential framework  for the target parameter $N$.
\subsection{Model assumptions}
A probabilistic model for the observed data relies on assumptions about the topology of the population graph and the sampling process.
\begin{assum}{\textbf{(Topology of the population graph).}}\label{assum:popgraph}
The population graph $G=(V,E)$ is an undirected, connected graph, with no self loops.
\end{assum}
For a given sampling procedure, this assumption implies that any vertex can reach any other vertex in the network, and that no vertex can sample itself. With the aim of leveraging information about the network to make statistical inference about the size of the population, one needs to make assumptions about the formation of the unobserved population graph. We can assume, for example, that the probability for any two vertices in the graph forming an edge is not affected by the presence of other vertices within the graph. This \textit{independence} assumption is critical for making valid inference given a sample from a networked population. If we set the probability for any two vertices sharing a tie to $\rho \in (0, 1)$, an homogeneous Erdos-Rényi random graph model $\mathcal{G}(N,\rho)$ arises \citep{erdds1959random}, where the parameter $\rho$ represents the \textit{density} of the network. Although simplistic given the potentially hidden and complex network structure in a hard-to-reach population, the Erdos-Rényi random model has been extensively used for population size estimation in link-tracing sampling and has been empirically proven to be useful in complex network settings \citep{crawford2018hidden}.
\begin{assum}{\textbf{(Random graph model).}}\label{assum:rgmodel}
The population graph is distributed according to the Erdôs-Rényi model: $G\sim \mathcal{G}(N,\rho)$.
\end{assum}
We now assume that the sampling process takes place within the population graph $G$ and progresses across vertices' ties. However, the process cannot be assumed known since the probability of any unit being recruited into the study may realistically depend on unobserved individual traits or network characteristics. Various authors have proposed approximations to similar designs under the assumption that the observed sample is a random realization from a design that is either independent of unobserved characteristics or entirely dependent on units' popularity (or degree), and that such information is accurately reported by participants (\citealt{Gile18,Vol08}).
\begin{assum}{\textbf{(Conditional independence under the Erdos-Rényi model).}}\label{assum:samprob}
For any unsampled vertex at any time of recruitment, the probability of being selected into the sample depends exclusively on edges it shares with recruiting vertices.
\end{assum}
Assumption \ref{assum:samprob} implies that an unrecruited unit's probability of appearing in the sample is not affected by edges shared with other unrecruited units at the time of recruitment. Additional assumptions regarding the sampling process and its conditional dependence on observable and/or unobservable individual traits will be discussed later.

Let $Y_{j(i)}=1$ if the $j$th unrecruited vertex shares an edge with the $i$th vertex at recruitment time $i$, where recruited individuals are ordered according to the time of their recruitment, and $Y_{j(i)}=0$ otherwise, with $j=1, \dots, N-i$. It is clear that the random variables $Y_{j(i)}$ are independent and identically distributed Bernouilli with probability $\mathbb{P}(Y_{j(i)}=1)=\rho$ given Assumption \ref{assum:samprob}. Let $\mathcal{E}_i$ denote the set of sampled vertices at recruitment time $i$. Then
\begin{equation}\label{eq:model1}
Y_i=\sum_{j\notin \mathcal{E}_i} Y_{j(i)} \sim \mbox{Binomial}(N-i,\rho),\,\, i=1,\dots, n.
\end{equation}
The random variable $Y_i$ takes values $y_i\in \{0, 1, \dots, N-i\}$ which are only observed given $G_U$, degree information of recruited individuals $\{d_i\}$, and recruitment times $\{t_i\}$. As $G_U$ cannot be fully observed given the study design, the rationale for dealing with this missing data problem under a frequentist setting is discussed next. 

When a unit is recruited into the study, they receive a fixed number $C$ of coupons to distribute among its unrecruited neighbours. For the $i$th sampled vertex, $Y_i<C$ implies that the number of coupons that they can give away will be less than $C$. This simple idea establishes a relationship between the topology of the augmented recruitment subgraph $G_U$ (thus $G$) and the number of coupons which were successfully distributed by each recruited unit. Let $C_i^{*}\in \{0, 1, \dots, C\}$ denote the number of coupons that the $i$th recruited unit distributed to their unrecruited neighbours, and define $y_i$ as a random realisation of $Y_i$.
\begin{assum}{\textbf{(Coupon distribution policy).}}\label{assum:givecoup1}
The topology of $G_U$ is linked to the recruitment process as follows:
\begin{itemize}
\item[(i)] If  $C^{*}_i=C$ then $y_i\geq C$.
\item[(ii)] If $C^*_i<C$ then $y_i=C^*_i$.
\end{itemize}
\end{assum}
Assumption \ref{assum:givecoup1} implies that (i) a recruiter who successfully distributes all $C$ coupons has at least $C$ unrecruited neighbours at the time of recruitment while (ii) recruiters who successfully distribute fewer than $C$ coupons have a corresponding network size of unrecruited neighbours equal to the number of coupons that they successfully distributed. This assumption is particularly realistic for cases in which there is a relationship of trust between a recruiter and its unrecruited neighbours, or when the incentive for participating in the study is highly lucrative, given that the recruiting unit has enough neighbours to recruit from, for a fixed value of $C$. In real-world RDS studies, a recruitment is effective when an unrecruited unit receives a coupon, returns it to an interviewer thereby confirming their interest to participate in the study. We make the additional assumption that a unit who voluntarily accepts a coupon will return it to the interviewer and be enrolled into the study.
\begin{assum}{\textbf{(Coupon acceptance policy).}}\label{assum:givecoup2}
Units who voluntarily accept a coupon proposal from a recruited neighbour become participants in the study.
\end{assum}
The plausibility of Assumption \ref{assum:givecoup2} could be driven by material incentives or a genuine interest in taking part in the study given its importance or benefits regarding issues that the target population faces. The case study in Section \ref{sec:casestudy} illustrates an RDS study conducted in Montréal (Canada), whose goals were to describe various aspects of sexual health and behaviors, and to understand the use of HIV prevention and care services as well as HIV and other Sexually Transmitted and Blood Borne Infections (STBBI) occurence within the GBM community. The majority of participants cited interest in issues affecting their community as their main reason for taking part in the study while $10\%$ cited monetary incentives \citep{lambert2019engage}. 
 
These assumptions imply that given the number of coupons distributed to new recruits, the corresponding number that will be returned depends on topological constraints imposed by $\{Y_i\}$, and is independent of units' decision-making or individual characteristics. This further establishes that the number of distributed coupons reveals partial information about the topology of $G_U$. In Section \ref{sec:simulation}, the performance of the approach described below is assessed under these assumptions via simulation; violations of Assumptions \ref{assum:givecoup1} and \ref{assum:givecoup2} will also be investigated.

\subsection{Model and likelihood}
Since $Y_i$ is not fully observed, we define random variables $Z_1, \dots, Z_n$ as
\begin{equation}
Z_i = \begin{cases}
Y_i &\text{if $Y_i< C$}\\
C &\text{if $Y_i \geq C.$}
\end{cases}
\end{equation}
Thus, random variables $\{Y_i\}$ are censored on the right and the transformed variables can be expressed as $Z_i=Y_i\wedge C$, $i=1, \dots, n$. For the $i$th individual, let $\delta_i=1$ if there is no censoring on the right, and $\delta_i=0$ otherwise. Thus, in addition to $G_T$ and $\{t_i\}$, we observe $(Z_i, \delta_i)$.

We now turn to the likelihood function for $(N, p)$. If $\delta_i=0$, the contribution to the likelihood for the $i$th individual is
\begin{eqnarray}
\nonumber \mathbb{P}[Z_i=z_i, \delta_i=0]&=&\mathbb{P}[Z_i=C|\delta_i=0]\mathbb{P}[\delta_i=0]\\
\nonumber &=&\mathbb{P}[\delta_i=0]\\
\nonumber &=&\mathbb{P}[Y_i\geq C].
 \end{eqnarray}
If $\delta_i=1$, then 
\begin{eqnarray}
\nonumber \mathbb{P}[Z_i=z_i, \delta_i=1]&=&\mathbb{P}[Y_i=z_i|\delta_i=1]\mathbb{P}[\delta_i=1]\\
\nonumber &=&\mathbb{P}[Y_i=z_i|Y_i< C]\mathbb{P}[Y_i< C]\\
\nonumber &=&\mathbb{P}[Y_i=z_i].
 \end{eqnarray}
Let $\bm{z}=(z_1,\dots, z_n)$ and $\bm{\delta}=(\delta_1, \dots, \delta_n)$ be $n\times 1$ ordered vectors according to recruitment times $\{t_i\}$. The likelihood for $(N, p)$ given $\bm{Z}$, $\bm{\delta}$, and $\bm{D}$ is given by
\begin{equation}\label{eq:Likelihood}
L(N,p; \bm{z}, \bm{\delta}, \bm{D})=\prod_{i=1}^n \left\lbrace  \mathbb{P}[Y_i=z_i]   \right\rbrace^{\delta_i}  \left\lbrace  \mathbb{P}[Y_i\geq C] \right\rbrace^{1-\delta_i},
\end{equation}
where
\begin{equation}
\mathbb{P}[Y_i=z_i]=\binom{N-i}{z_i} \rho^{z_i}(1-\rho)^{N-i-z_i}
\end{equation}
and 
\begin{equation}
\mathbb{P}[Y_i\geq C]=1-\sum_{k=0}^{C-1} \binom{N-i}{k} \rho^k(1-\rho)^{N-i-k}.
\end{equation}
The maximization of the likelihood function (\ref{eq:Likelihood}) is straightforward in most programming languages for statistical computing. To prove consistency and asymptotic normality for the maximum likelihood estimator $(\hat N, \hat \rho)$ of $(N, \rho)$, we derive a new likelihood function by working with the conditional distribution of $\bm{Z}$ given $R=\sum_{i=1}^ n \delta_i$. Let $\bm{X}=(X_1, \dots, X_R)$ be a vector representing the elements of $\bm{Z}$ for which $Z_i=Y_i$.  Since $\mathbb{P}[Z_i=C|\delta_i=0]=1$, the conditional likelihood $L_C$ for $(N, p)$ given $\bm{X}=\bm{x}$ and $R=r$ is given by
\begin{equation}\label{eq:LikelihoodC}
L_C(N, \rho; \bm{x}, r)=\prod_{i=1}^r \frac{\mathbb{P}[X_i=x_i]}{\mathbb{P}[X_i< C]}.
\end{equation}
Let $h$ be the solution to
\begin{equation}\label{eq:solveMLE}
L_C(h, \hat{\rho}(h); \bm{x}, r)=L_C(h-1, \hat{\rho}(h-1); \bm{x}, r),
\end{equation}
where $\hat{\rho}(h)$ is the solution to $\partial \log \{L_C(h, \rho; \bm{x}, r)\}/\partial \rho=0$. The MLE is given by $\hat{N}=[h]$, where $[x]$ represents the greatest integer less than $x$. Note that for $\rho$ unknown, the MLE $\hat{N}$ may be infinite. In fact, consider the non-truncated case for which the moment estimator for $N$ is $\tilde{N}=\bar{Y}/\tilde{\rho}+(n+1)/2$, where $\tilde{\rho}=1-S^2/\bar{Y}^2$, $\bar{Y}=(1/n)\sum_{i=1}^n Y_i$ and $S^2=(1/n)\sum_{i=1}^n (Y_i-\bar{Y})^2$. This clearly leads to a negative estimate for $\rho$ if $S^2>\bar{Y}^2$, which would thus be set to 0 and therefore lead to an infinite estimate of N.

\subsection{Inferential procedures for $N$}
This section examines asymptotic properties for the MLE of $N$. We develop the inferential framework by assuming $\rho$ known before proposing an extension when both $N$ and $\rho$ are to be estimated. All proofs, as well as  consistency and asymptotic normality results for when $\rho$ is unknown, are deferred to the Online Supplementary Material (Web Appendix A).
\begin{prop}{\textbf{(Consistency of the MLE $\hat N$).}}\label{defi:consistencyN}
For $N\rightarrow \infty$, $n$ (and $r$) fixed or increasing such that $N-n\rightarrow \infty$, and for $\rho$ known,
$$
\lim_{N\to\infty} \mathbb{P}[\hat N=N]=1.
$$
\end{prop}
The next theorem establishes asymptotic normality for the MLE $\hat{N}$.
\begin{theor}{\textbf{(Asymptotic normality for the MLE $\hat N$).}}\label{defi:normalityN}
For $N\rightarrow \infty$, $n, r \rightarrow \infty$ such that $r/(N-r)\rightarrow 0$, and for $\rho$ known,
\begin{equation}\label{eq:asymptoticnorm}
\left[ \frac{\sum_i\sqrt{N}\rho/\{(N-i)(1-\rho)\}-Q(N,\rho) }{\sqrt{\sum_i N\rho/\{(N-i)(1-\rho)\} } }  \right] \left(\hat{N}-N\right) \xrightarrow[]{\mathcal{D}} N\left(0, 1\right),
\end{equation}
and
\begin{eqnarray}
\nonumber Q(N, \rho)&=&\frac{\sum_{k=0}^{C-1} \binom{N-r-1}{k} \rho^k(1-\rho)^{N-r-1-k}G(N-r-1;N-k-r-1)}{\sum_{k=0}^{C-1} \binom{N-r-1}{k} \rho^k(1-\rho)^{N-r-1-k}}\\
\nonumber &-&\frac{\sum_{k=0}^{C-1} \binom{N-1}{k} \rho^k(1-\rho)^{N-1-k}G(N-1;N-k-1)}{\sum_{k=0}^{C-1} \binom{N-1}{k} \rho^k(1-\rho)^{N-1-k}},
\end{eqnarray}
where
$$
G(N-1;N-k-1)=\sum_{j=1}^{N-1} \frac{1}{j}-\sum_{j=1}^{N-k-1} \frac{1}{j}=\sum_{j=N-k}^{N-1}\frac{1}{j}.
$$
\end{theor}


Having established an estimator for $N$ that is satisfied under somewhat rigid assumptions regarding the coupon distribution, we next turn to the derivation of bounds that can be used when those assumptions do not hold.
\section{Violations of the coupon distribution policy}\label{sec:identbound}
First, we present the terminology for characterizing the nature of violations of Assumptions \ref{assum:givecoup1} and \ref{assum:givecoup2}. We use the term \textit{practical violation} to describe violations that can be established from the observed data; the term \textit{hypothetical violation} describes violations that cannot be described nor tested using the observed data.

The coupon distribution policy (Assumption \ref{assum:givecoup1}) is violated when a unit successfully distributes fewer coupons than what would have been expected given the number of unrecruited neighbours at the time of that unit's recruitment. A clear violation of this policy can be easily established for the $i$th participant, for example, when $d_i-(i-1)\geq C$ and $C^*_i<C$; this is a case of practical violation. The random variable $Y_i$, in this instance, takes values in  the set $\{\max(d_i-i+1, C^*_i), \max(d_i-i+1 C^*_i)+1, \dots, d_i-1\}$. A hypothetical violation of Assumption \ref{assum:givecoup1} implies that $Y_i$ takes values in $\{C^*_i, C^*_i+1,\dots, d_i-1\}$. Deriving a consistent point estimator for $N$ without an informative probability distribution on these sets is challenging, if not impossible. Section \ref{sec:sensitivity} investigates the sensitivity of the proposed inferential framework when random realizations of $Y_i$ are set to $y_i=C^*_i$ if $C^*_i<C$.

Now consider the coupon acceptance policy (Assumption \ref{assum:givecoup2}). Let $d^T_i=\sum_{j\in V_T}\mathbbm{1}\{(i,j)\}-1$ denote the number of recruits by the $i$th sampled unit; Assumption \ref{assum:givecoup2} is violated if $d^T_i\neq C^*_i$. This represents a case of hypothetical violation as $C^*_i$ is usually not reported in classical RDS studies, and will impact the assessment of the coupon distribution policy if Assumption \ref{assum:givecoup1} is also violated. The sensitivity analysis of Section \ref{sec:sensitivity} considers cases of simultaneous violations of both assumptions.

Violations of the coupon distribution and acceptance policies imply that random realizations of $\{Y_i\}$ are not observed, inducing a missing data problem that affects the validity of statistical inference for $N$. To formalize this problem, we introduce the notion of \textit{concordancy} for hypothetical realizations $\{y_i\}$ of $\{Y_i\}$.
\begin{defi}{\textbf{(Concordancy).}}\label{def:Concord}
The dataset $\tilde{D}=\{\tilde{y}_i; i\in V_T\}$ is concordant with $\{(y_i, d_i); i\in V_T\}$ if $C^*_i\leq \tilde{y}_i\leq d_i-1$, $i=1,\dots, n$.
\end{defi}
Let $\mathcal{F}(\bm{y},\bm{d})$ denote the set of data that are concordant with $\{(y_i, d_i); i\in V_T\}$, where $\bm{y}=(y_1,\dots, y_n)$ and $\bm{d}=(d_1,\dots, d_n)$. Since we can construct two sets of concordant data that will give rise to the same distribution for the observed data, the target parameter $N$ is not identifiable. Next, we propose an algorithm for sampling from the concordancy space $\mathcal{F}(\bm{y},\bm{d})$ and deriving lower and upper limits for the smallest interval that contains $N$.

\subsection{Sampling from the concordancy space $\mathcal{F}(\bm{y},\bm{d})$}\label{sec:subspacesampl}
We explore the set $\mathcal{F}(\bm{y},\bm{d})$ by adding and substracting a unit from each $\tilde{y}_i$, $i=1,\dots, n$. Let $\tilde{D}^{(0)}=\{\tilde{y}^{(0)}_i; i\in V_T\}\in \mathcal{F}(\bm{y},\bm{d})$ denote the initial concordant data proposal. Let $\tilde{d}_i=\max(d_i-i+1,C^*_i)$;  we can set, for example, $\tilde{y}^{(0)}_i=\tilde{d}_i$ or $\tilde{y}^{(0)}_i=d_i-1$, $i=1, \dots, n$. The next concordant dataset $\tilde{D}^{(1)}$ is obtained according to the following procedure. We randomly sample $i\in V_T$. If $\tilde{y}^{(0)}_i=\tilde{d}_i$, then compute $\tilde{y}^{(1)}_i=\tilde{y}^{(0)}_i+1$; if $\tilde{y}^{(0)}_i=d_i-1$, then compute $\tilde{y}^{(1)}_i=\tilde{y}^{(0)}_i-1$; finally, if $\tilde{d}_i<\tilde{y}^{(0)}_i<d_i-1$, then add or substract the unit value from $\tilde{y}^{(0)}_i$ according to a uniform distribution. This is summarized in Algorithm \ref{alg:recAlgo}.

\begin{algorithm}\label{alg:recAlgo}
{\small
\caption{The algorithm for sampling from $\mathcal{F}(\bm{y},\bm{d})$}
\KwData{Start with the current concordant data $\tilde{D}=\{\tilde{y}_i; i\in V_T\}$}
$\tilde{D}^{new}\gets \tilde{D}$\;
Randomly sample $i\in V_T$\;
  \eIf{$\tilde{y}_i=\tilde{d}_i$ }{
   $\tilde{y}^{new}_i\gets \tilde{y}_i+1$\;
  }{\eIf{$\tilde{y}_i<d_i-1$ and $\tilde{y}_i>\tilde{d}_i$}{
   $A\sim \mbox{Unif}(\{-1,0,1\})$; $\tilde{y}^{new}_i\gets \tilde{y}_i+A$ 
    }{$\tilde{y}^{new}_i\gets \tilde{y}_i-1$}
}
}
\end{algorithm}

\subsection{Derivation of the identification region for $N$}\label{sec:stocop}
We present a stochastic optimization technique for obtaining the (global) maximum and minimum of the population size $N$. This approach relies on Algorithm \ref{alg:recAlgo} for efficiently exploring the concordancy space $\mathcal{F}(\bm{y},\bm{d})$ and a variant of the quadratic programming technique \citep{de2014identification} for optimizing a monotonic function of $N$. We adopt the strategy presented by \cite{crawford2018identification} in which a function of a graph attribute is defined in such a way that its maximum coincides with an optimum for the target graph parameter.

Let $N(\tilde{D})$ represent an estimate for $N$ given a concordant dataset $\tilde{D} \in \mathcal{F}(\bm{y},\bm{d})$; let $S(x)$ denote a function taking $N(\tilde{D})$ as argument, where $x\in \{n, n+1, \dots, N_0\}$, where $N_0$ represents the maximum value of $N$. Let $R(N)\propto \exp\{S(N)\}$ represent the objective function to be maximized for sets $\tilde{D} \in \mathcal{F}(\bm{y},\bm{d})$. We construct an MCMC algorithm that takes $\tilde{D}^{(t)}$ at step $t$ and accept a new proposal $\tilde{D}^{(t+1)}$ according to the procedure described in Algorithm \ref{alg:recAlgo} with probability 
$$
\beta_t=\min\left\lbrace 1, \exp\left[\frac{R\{N(\tilde{D}^{(t+1)})\}-R\{N(\tilde{D}^{(t)})\}}{B_t}\right]\right\rbrace,
$$
where $B_t$ is a non-increasing sequence taking positive values, and $\lim_{t\to\infty} B_t=0$. Let $S(N)=\{\epsilon+(N-n)/n^{\nu}\}^{-1}$ for $\epsilon>0$ and $\nu\geq 1$; note that the maximum of $S(N)$ coincides with the lower bound of $N$. The next proposition establishes the convergence of $S(N(\tilde{D}))$ to a global maximum given a specification of the sequence $B_t$, $t=1,2,\dots$.
\begin{prop}{\textbf{(Convergence of the algorithm).}}\label{prop:conv}
For $\epsilon>0$, let $B_t=\{\epsilon\log(t)\}^{-1}$. Let $\mathcal{R}$ denote the set of concordant data $\tilde{D} \in \mathcal{F}(\bm{y},\bm{d})$ for which the objective function $R(N(\tilde{D}))$ reaches its global maximum. Then
\begin{equation}\label{eq:conv}
\lim_{t\to\infty} \mathbb{P}\left[\tilde{D}^{(t)} \in \mathcal{R}\right]=1.
\end{equation}
\end{prop}
The proof is presented in the online appendix. To derive the upper bound, a candidate function is $S(N)=\{\epsilon+(N_0-N)/N_0^{\nu}\}^{-1}$ for $\nu\geq 1/2$. Note that the resulting identification bounds are sharp. To boost the speed of the algorithm, one can start with the proposal $\bm{\tilde{y}}=(d_1-1,\dots, d_n-1)$ when the goal is the minimization of the objective function to derive the upper bound of $N$ since $\mathbb{P}[Y_i\geq x]\geq \mathbb{P}[Y_{i+1}\geq x]$ for all $x\in \mathbb{N}$. Similarly, the starting point of $\bm{\tilde{y}}=(\tilde{d}_1,\dots, \tilde{d}_n)$ can be used to efficiently minimize the objective function to determine the lower bound of $N$.
\section{Assessment using simulated RDS data }\label{sec:simulation}
The goal of this study is two-fold: to (i) evaluate the inferential framework presented in Section \ref{sec:methodo} and, in particular, to assess the accuracy of the MLE for $\hat N$, the coverage for the corresponding 95\% (Wald-type) confidence interval of $N$ given the observed data, and (ii) conduct investigations into the robustness of the procedure when Assumptions \ref{assum:givecoup1} and \ref{assum:givecoup2} are violated.  We then turn to an evaluation of the proposed method for bounding $N$ in settings where the MLE is not consistent due to the violation of Assumptions \ref{assum:givecoup1} and \ref{assum:givecoup2}. A comparative study with the (Bayesian) successive sampling approach of \cite{handcock2015estimating} is presented in the Online Supplementary Material (Web Appendix B); other competing approaches were not considered because their implementation requires multiple RDS surveys or external information (e.g., prevalence of a trait in the population for the multiplier method).

\subsection{Evaluation of the proposed population size estimator and inferential framework}
Random realisations $y_i$ of $Y_i$, $i=1, \dots, n$, were generated for population sizes $N=5\times 10^3$, $1\times 10^4$, with corresponding sample fractions $n/N=2.5\%,\,5\%$, and network density $\rho=1\%$, under the homogeneous Erdôs-Rényi model. The number of coupons $C$ is set as follows. Let $q_y(\alpha)$ denote the $\alpha$-th quantile of $\{y_i\}$, i.e., the smallest integer for which $\mathbb{P}[Y_i\leq q_y(\alpha)]\geq \alpha$. We set $C=q_y(\alpha)$, meaning that $n\times (1-\alpha)$ random observations will be censored on the right, for $\alpha=25\%, 50\%$.

The maximization of the likelihood function (\ref{eq:Likelihood}) was achieved using a \cite{nelder1965simplex} optimizer; its implementation requires starting values for $(N, \rho)$ which were obtained as follows. First, we expressed the likelihood for $(N, \rho)$ given the uncensored observations $U=\{y_i: y_i<C\}$ as a single function of $N$,
\begin{equation}\label{eq:Likelihood2}
L(N; U)=\prod_{i=1}^n \binom{N-i}{\delta_iy_i} \left\lbrace \frac{\sum_{i=1}^n \delta_iy_i}{\sum_{i=1}^n (N-i)} \right\rbrace^{\delta_iy_i} \left\lbrace 1-\frac{\sum_{i=1}^n \delta_iy_i}{\sum_{i=1}^n (N-i)} \right\rbrace^{N-i-\delta_iy_i}.
\end{equation}
We then maximized (\ref{eq:Likelihood2}) to obtain the starting value for $N$, $N_{start}$, and deduced the corresponding estimate for $p$ as $p_{start}=\sum_{i=1}^n \delta_iy_i/\{\sum_{i=1}^n (N_{start}-i)\}$.


\begin{table}
\caption{\small Relative bias (RB) for $\hat N$, relative root mean squared error (RRMSE) for $\hat N$, coverage for the 95\% confidence interval of $N$ (95\% Cov.)  and its corresponding relative length (RLCI), for increasing population size ($N$), coupon percentile ($\alpha$), and sample size $n$.}
\begin{center}
\setlength\extrarowheight{-11.5pt}
\small
\begin{tabular}{lllccccccc} \toprule
{$N$} & {$n$} & &{$\alpha$} & {$RB(\hat N)$}  & {$RRMSE$ }&{$95\%$ Cov. }& {$RLCI$ }\\ \toprule
\multirow{8}{1em}{$5\times 10^3$} & \multirow{3}{1em}{$500$}  & & $25\%$     & 0 & 5.33& 0.98 & 0.20 \\
& \multirow{12}{1em}{$1000$}   && $50\%$   & 0&  4.06 &  0.96 & 0.11 \\
&   && $75\%$     & 0&  2.26 &  0.97 &  0.09 \\
&   &&      & &      && \\
&   && $25\%$     & 0 & 2.20 & 0.98 & 0.12 \\
&   & & $50\%$    & 0 & 1.62 & 0.97 &  0.09 \\
&    && $75\%$     & 0 & 1.09 &  0.95 & 0.06 \\
\hline 
\multirow{8}{1em}{$1\times 10^4$} & \multirow{3}{1em}{$500$}  & & $25\%$     &0 & 5.17 & 0.97 & 0.14 \\
& \multirow{12}{1em}{$1000$}   && $50\%$   & 0  & 3.34 & 0.95 & 0.08 \\
&   && $75\%$     & 0& 2.19 & 0.96 & 0.06 \\
&   &&      & &      && \\
&   && $25\%$     & 0 & 2.59 & 0.98 & 0.09 \\
&   & & $50\%$    & 0 & 1.41 & 0.98 & 0.06 \\
&    && $75\%$     & 0  & 1.23 & 0.96 & 0.05 \\
\bottomrule
\end{tabular}
\end{center}
\label{table:simulation_CVGE}
\end{table}

For $B=1000$ random samples, we computed the relative bias and the relative root mean squared error of $\hat{N}$ as 
$$
RB(\hat N)=\frac{B^{-1}\sum_{b=1}^B \hat{N}_b- N}{N},\,\, RRMSE(\hat N)=\frac{MSE(\hat N)^{1/2}}{N},
$$
where $\hat{N}_b$ is the estimate from the $b$th repetition, and $MSE(\hat N)=\sum_b (\hat{N}_{b}-N)^2/B$. 
For each replication $b$ of the Monte Carlo simulation, we computed an estimate for the asymptotic variance of $\hat N$, $v(\hat N)$.
The $95\%$ confidence interval for $\hat N$ is calculated as $\exp[\log(\hat N) \pm 1.96\,\mbox{s.e.} \{\log(\hat N)\}]$, where $\mbox{s.e.} \{\log(\hat N)\}\approx \{v(\hat N)\}^{1/2}/\hat{N}$; the expected relative length of the 95\% confidence interval is computed as $RLCI(\hat N)=(UB-LB)/N$, where $UB$ and $LB$ are the expected corresponding upper and lower bounds, respectively.

The results are presented in Table \ref{table:simulation_CVGE}. The MLE for $N$ is unbiased across all simulation scenarios, with a decreasing relative root mean squared error as each of $N$, $n$, and $\alpha$ increase. The coverage of the 95\% confidence interval for $N$ is greater than or equal to its nominal value; it grows closer to 95\% with increasing $N$, $n$, or $\alpha$.
\subsection{Robustness of the inferential procedure}\label{sec:sensitivity}
This section investigates the robustness of our inferential procedure when Assumption \ref{assum:givecoup2} is relaxed. For a given recruiter, let $\lambda\in (0, 1)$ represent the proportion of unrecruited neighbours who are not available for coupon distribution; let $\eta$ denote the proportion of recruiters for whom $\lambda\neq 0$. We conduct a study to investigate the impact of increasingly important violations of the assumption, as controlled by setting $\lambda=5\%,\, 10\%,\, 25\%,\, 50\%$ and $\eta=10\%,\, 25\%$; we set $N=5\times 10^3$, $\alpha=25\%$ and $n=500$.

\begin{table}
\caption{\small Relative bias (RB) for $\hat N$, and relative root mean squared error (RRMSE) for $\hat N$, coverage for the 95\% confidence interval of $N$ (95\% Cov.) and its corresponding relative length (RLCI).}
\begin{center}
\setlength\extrarowheight{-11.5pt}
\begin{tabular}{lllccccccc} \toprule
{$N$} & {$\eta$} & &{$\lambda$} & {$RB(\hat N)$}  & {$RRMSE$ }&{$95\%$ Cov. }& {$RLCI$ }\\ \toprule
\multirow{8}{1em}{$5\times 10^3$} & \multirow{3}{1em}{$10\%$}  & & $5\%$     & -0.03  &10.03  &0.97  &0.18\\
& \multirow{11}{1em}{$25\%$}   && $10\%$   & -0.03  &9.96 &0.96  &0.18\\
&   && $25\%$     & -0.03  & 9.46 & 0.96&  0.18 \\
&   && $50\%$     &-0.09  &44.15 &0.40 & 0.16\\
&   &&      & &      && \\
&   && $5\%$     & -0.07  &32.39 & 0.64 & 0.17 \\
&   & & $10\%$    & -0.07  &33.45 & 0.62 & 0.16 \\
&    && $25\%$     & -0.08  &33.66 & 0.62 & 0.17 \\
&   && $50\%$     & -0.20  & 191.88 &  0.10 &  0.13\\
\bottomrule
\end{tabular}
\end{center}
\label{table:simulation_sens}
\end{table}

Table \ref{table:simulation_sens} presents the performance of the estimator and inferential procedure for $N$. The estimator $\hat N$ is nearly unbiased across all sample sizes, with a 96\% coverage rate for $\lambda=5\%, 10\%,\,25\%$ when $\eta=10\%$. However, the estimator exhibits substantial bias when $\lambda=50\%$ and $\eta=25\%$, with a 10\% coverage rate and a substantially bigger standard error. This shows that the proposed methodology is robust to moderate deviations from Assumptions \ref{assum:givecoup1} and \ref{assum:givecoup2} in terms of bias, however the inferential framework does not recover the true uncertainty associated with the population size estimator under modest deviations from model assumptions.

\subsection{Finding the identification bounds of $N$}
\begin{figure}
  \centering
  \begin{minipage}[b]{0.45\textwidth}
    \includegraphics[width=\textwidth]{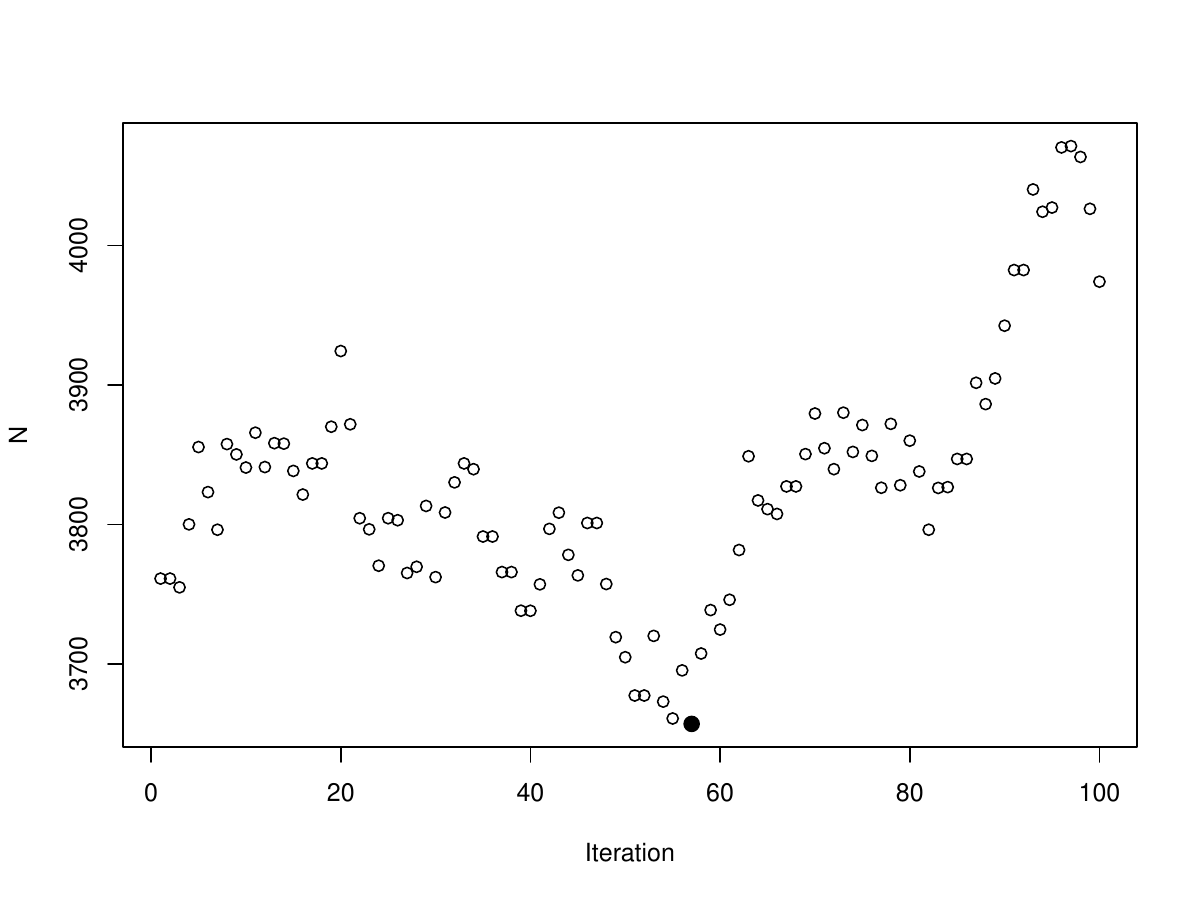}
  \end{minipage}
  \begin{minipage}[b]{0.45\textwidth}
    \includegraphics[width=\textwidth]{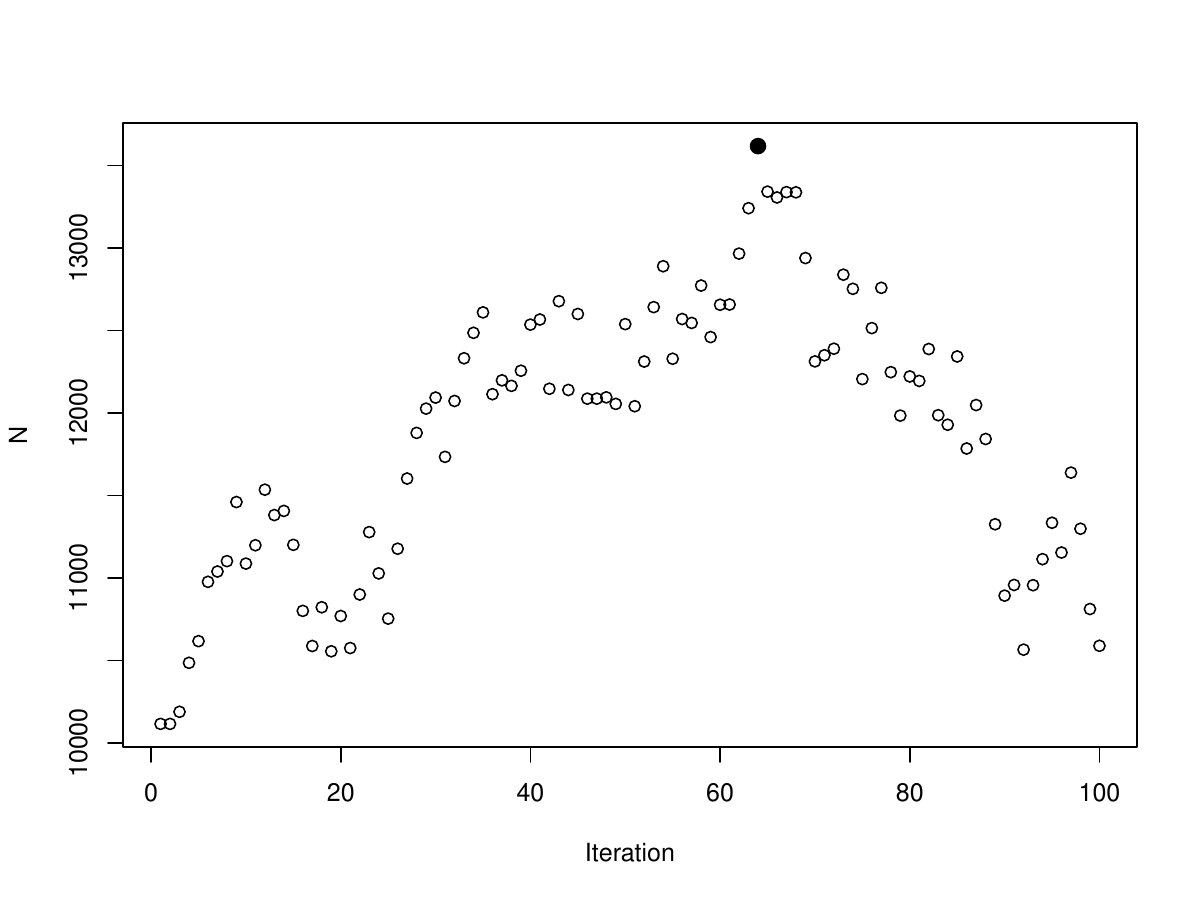}
  \end{minipage}
    \caption{{ \small Illustration of the algorithm for obtaining the identification region of the population size $N$ at each iteration. On the left (on the right), the iterations for obtaining the lower (upper) bound of $N$, where the black circle represents the global minimum (maximum). The corresponding identification region is [3657; 13618] for $N=5000$, $n=500$, and $\rho=1\%$.}}
\label{fig:bounds}
\end{figure}

This section implements the stochastic optimization algorithm presented in Section \ref{sec:stocop} for obtaining the identification region of the target parameter $N$, an approach necessitated when Assumptions \ref{assum:givecoup1} and \ref{assum:givecoup2} are violated so that a consistent estimator of $N$ is no longer guaranteed. In the first setting, realizations $y_i$ of $Y_i$, $i=1, \dots, n$ are drawn for $N=5000$, $n=500$, and $\rho=1\%$. To find the lower bound, we set $\epsilon=2.2$, $\delta=3/2$, and used the starting proposal $\bm{\tilde{y}}=(\tilde{d}_1,\dots, \tilde{d}_n)$. The upper bound was derived by setting $\epsilon=1.2$, $\delta=1/2$, $N_0=1\times 10^5$, with the starting proposal $\bm{\tilde{y}}=(d_1-1,\dots, d_n-1)$. The results are presented in Figure \ref{fig:bounds}.

An extensive investigation into the performance of the algorithm for a range of population and sample parameters, with a particular focus on the choice of values for $\epsilon$ and $\delta$, is presented in the Online Supplementary Material (Web Appendix C).

\section{Application to real-word RDS datasets}\label{sec:casestudy}
This section analyzes two RDS datasets. The first studies people who inject drugs, sometimes termed injection drug users (IDU), in the Kohtla-Järve region of Estonia \citep{wu2017using}. The second studies the GBM community in the Metropolitan area of Montreal \citep{lambert2019engage}, with the goal of infering $N$ for the respective target populations.

\subsection{IDU in the Kohtla-Järve region of Estonia}
The study was conducted from May to July of 2012 and targeted the IDU population in the Kohtla-Järve region of Estonia \citep{burke2015tale}. Eligible participants were 18+ years old individuals who injected drugs in the four weeks prior to the study and who spoke Russian or Estonian. The recruitment process started with the recruitment of six seeds; these seeds and all recruits received three coupons for peer recruitment. A total of $n=600$ participants took part in the study across 11 recruitment waves. 
\begin{figure}
\begin{center}
\includegraphics[scale=.45]{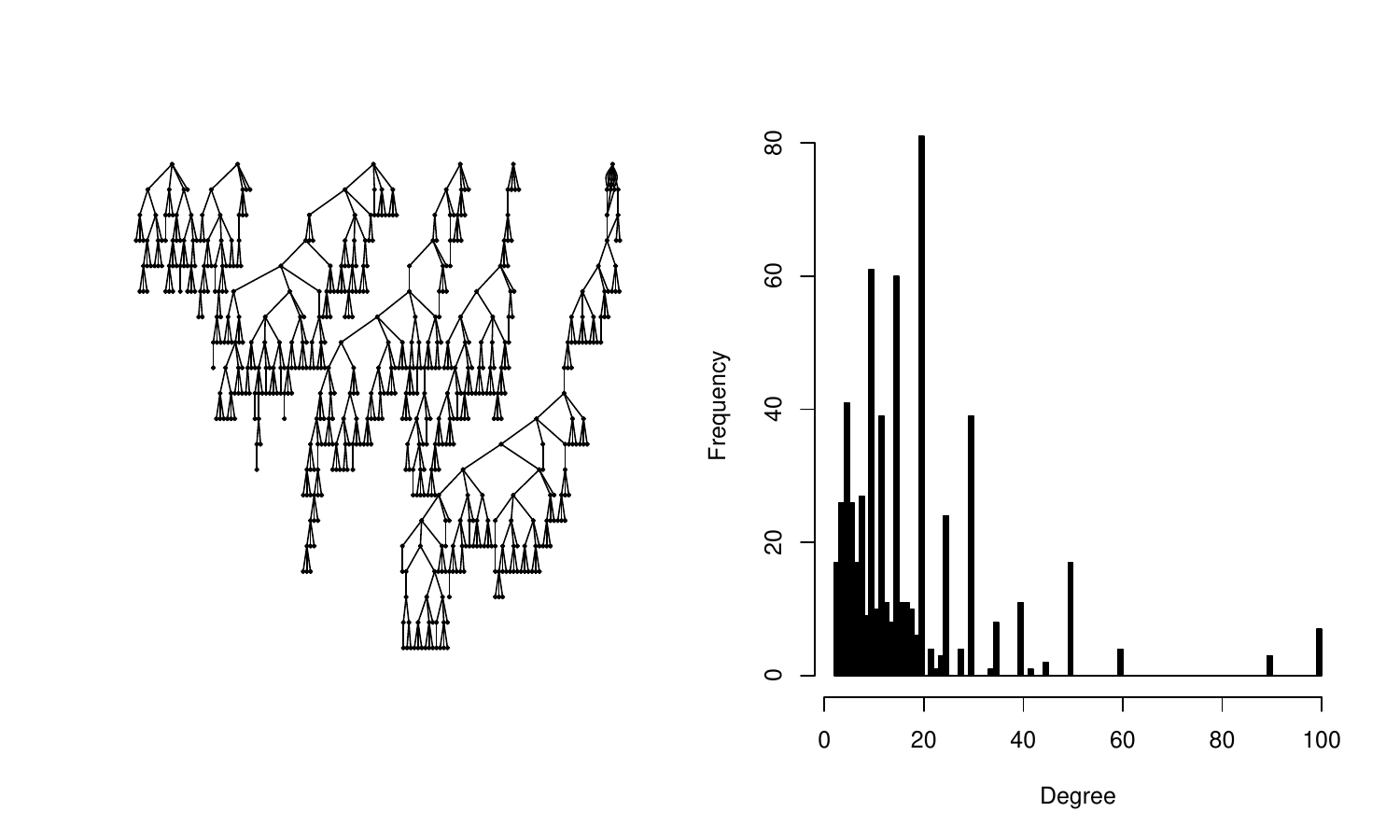}
\end{center}
\vspace*{-0.5cm}
\caption{ Graphical representation of the RDS dataset collected from a sample of the IDU population in the Kohtla-Järve region of Estonia. On the left, the recruitment tree; on the right, the degree distribution.}
\label{fig:Estdata}
\end{figure}

The recruitment data are displayed in Figure \ref{fig:Estdata}. The average degree is 17 (with a standard deviation of 15.10) and the median degree is 15. Approximately 61\% of participants did not recruit; 12\% of participants recruited one or two peers while 27\% successfully recruited 3 participants. Descriptive statistics show that 3 participants recruited fewer participants than the mininum of unrecruited neighbours who were available for recruitment. This constitutes a violation of the coupon distribution policy stated in Assumption \ref{assum:givecoup1}; these cases account for only 0.5\% of the sample, so the proposed methodology will likely provide valid point inference for $N$.

Fitting the model to the observed data yields $\hat{N}=744$ with a 95\% confidence interval of $[710; 779]$. The same dataset was analyzed by \cite{wu2017using}; they obtained an estimated population size of $654$ (95\% CI of 509-804) for the multiplier method, while 
the successive sampling approach of \cite{handcock2015estimating} gave point estimates between 600 and 2500. The (Bayesian) network-based approach of \cite{crawford2018hidden}, conditional on priors for $N$ and $\rho$, yielded a posterior mean of approximately 2000 (95\% posterior credible interval of 1700-2500), yielding an interval that is more than 30\% larger than that using the methodology proposed here.

Applying the optimization algorithm gives lower and upper bounds of 623 and 2204 for the IDU population in the Kohtla-Järve region; the corresponding bounds for the network-based approach  \citep{crawford2018hidden} are 700 and 2800, respectively.

\subsection{Estimating the size of the GBM community in Montréal}
We have access to an RDS dataset from the Engage study, a cohort study that took place in Montréal, Toronto, and Vancouver, with the goal of providing an accurate picture of the overall sexual health and behaviors of the GBM community. Individuals eligible for participation were French or English-speaking, cis- or transgender men 16+ years old, who reported sex with at least one man in the six months prior to study visit, and who resided in Vancouver, Toronto, or Montreal. Participants completed a biobehavioral questionnaire, and also completed testing for a variety of sexually transmitted infections including HIV, and were subsequently invited to take part in follow up visits (\citealt{lambert2019engage, cox2021use}). Only those participants from Montreal are included in this analysis.

\begin{figure}
\begin{center}
\includegraphics[scale=.45]{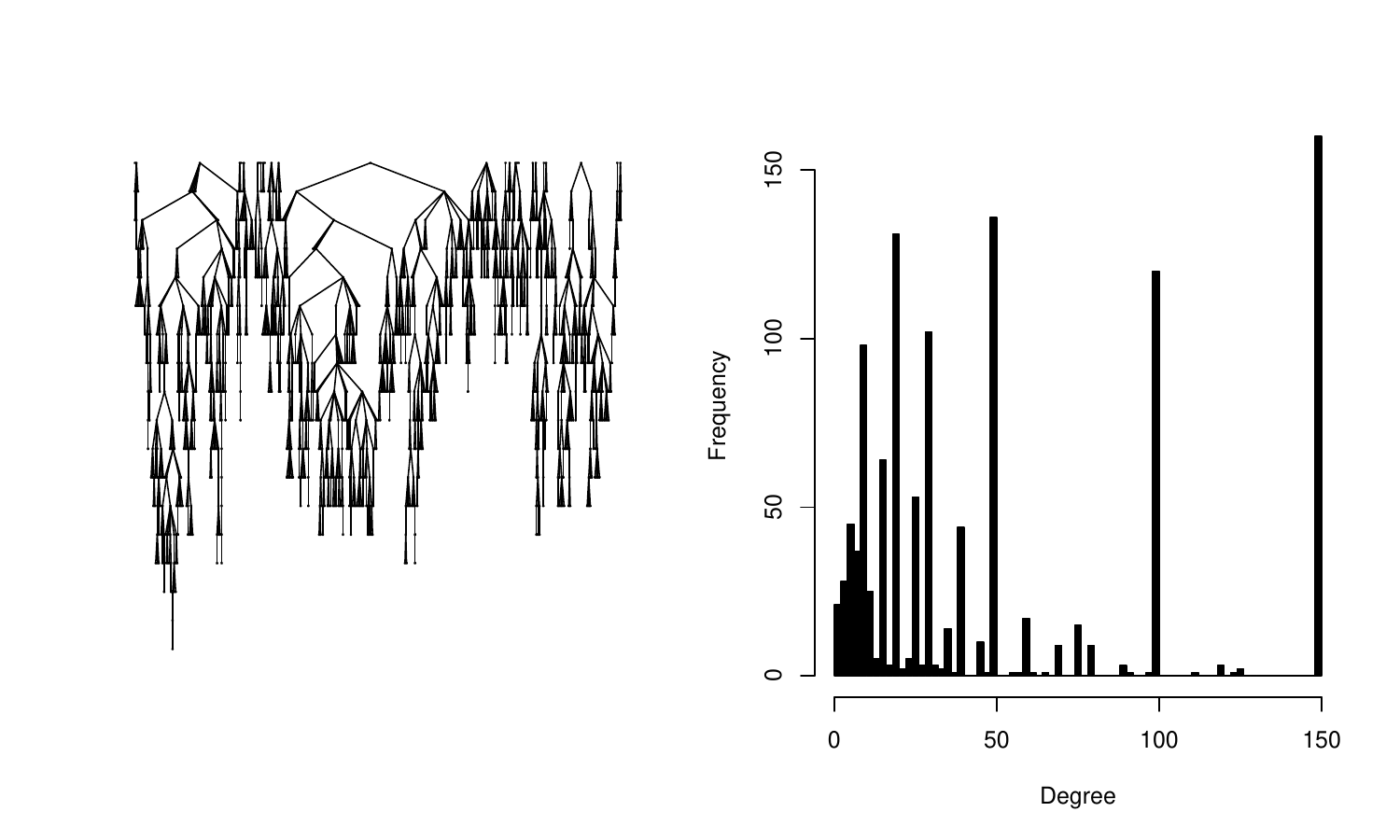}
\end{center}
\vspace*{-1.2cm}
\caption{{ Graphical representation of the RDS dataset collected from a sample of $n=1179$ gay, bisexual and other men who have sex with men (GBM) in in the Montréal. On the left, the recruitment tree; on the right, the degree distribution}}
\label{fig:engagenet}
\end{figure}

The recruitment process was conducted as follows. Twenty-seven (27) members were initially selected as seeds following a formative assessment and community mapping. The initial seeds identified as French Canadian (17), English Canadian (1), European (4), Arab (1), South-East Asian (1), and mixed (2); four seed participants were living with HIV. Each recruit received six (6) coupons to distribute among their personal network of contacts; this number, according to surveyers, was deemed appropriate given the target sample size of approximately 1200 GBM members. A monetary reward of \$50 was given to each participant who completed the questionnaire and underwent testing; an additional \$15 was awarded for each additional recruit who presented to the study site, for a maximum of six. A total of $n=1179$ members of the GBM community were recruited from February 2017 through June 2018. The observed RDS recruitment graph is illustrated in Figure \ref{fig:engagenet}.

Approximately 45\% of participants who were given coupons recruited at least one member; six seeds were not able to recruit. Among those who successfully recruited at least one participant, 82\% brought in between one and three members; the median for the number of recruits is two. Around 99\% of participants declared having received a coupon from someone they know as a friend or sexual partner. Summary statistics for the total number of recruits by each participant show that 1.5\% (18 participants) of recruiters successfully brought six participants into the study, implying that there is censoring on the right for the corresponding observations.

\begin{figure}
  \centering
  \begin{minipage}[b]{0.48\textwidth}
    \includegraphics[width=\textwidth]{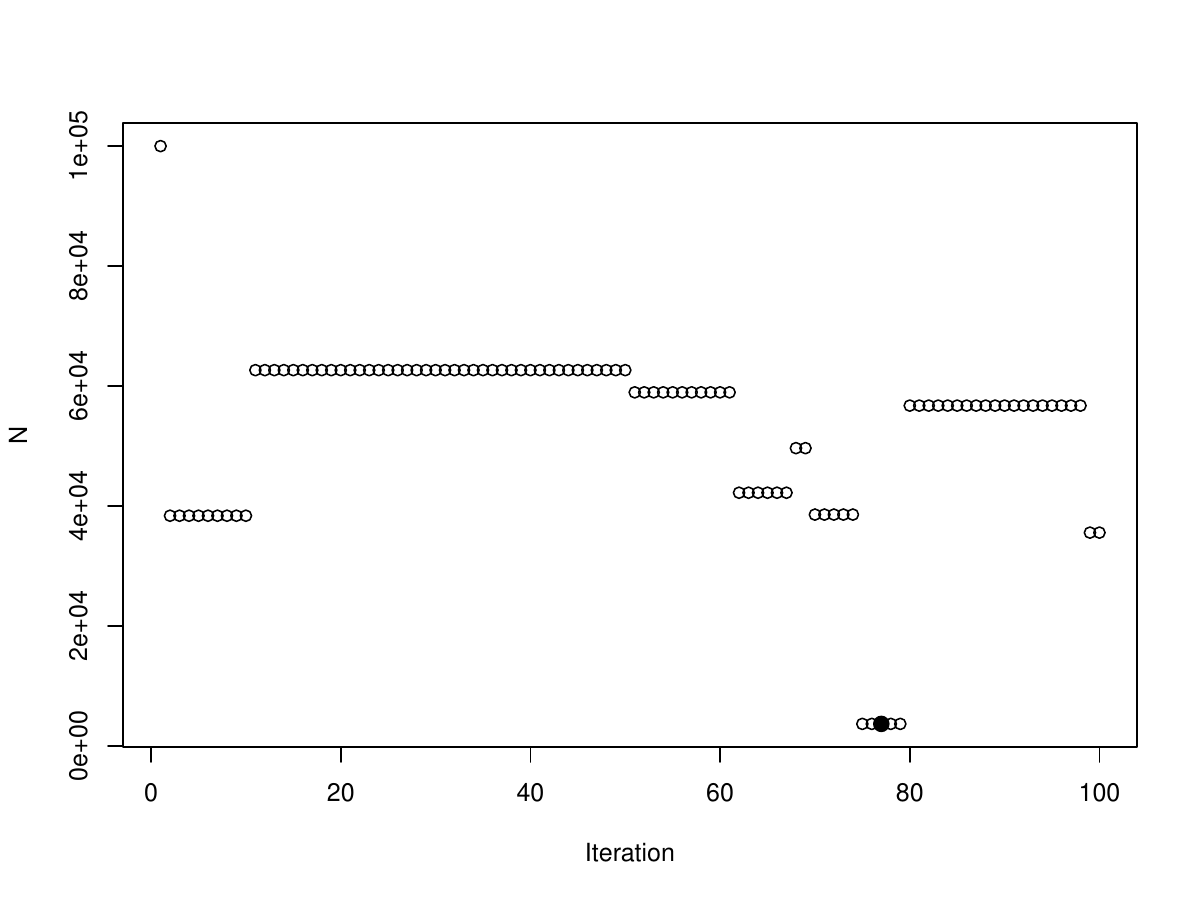}
  \end{minipage}
  \begin{minipage}[b]{0.48\textwidth}
    \includegraphics[width=\textwidth]{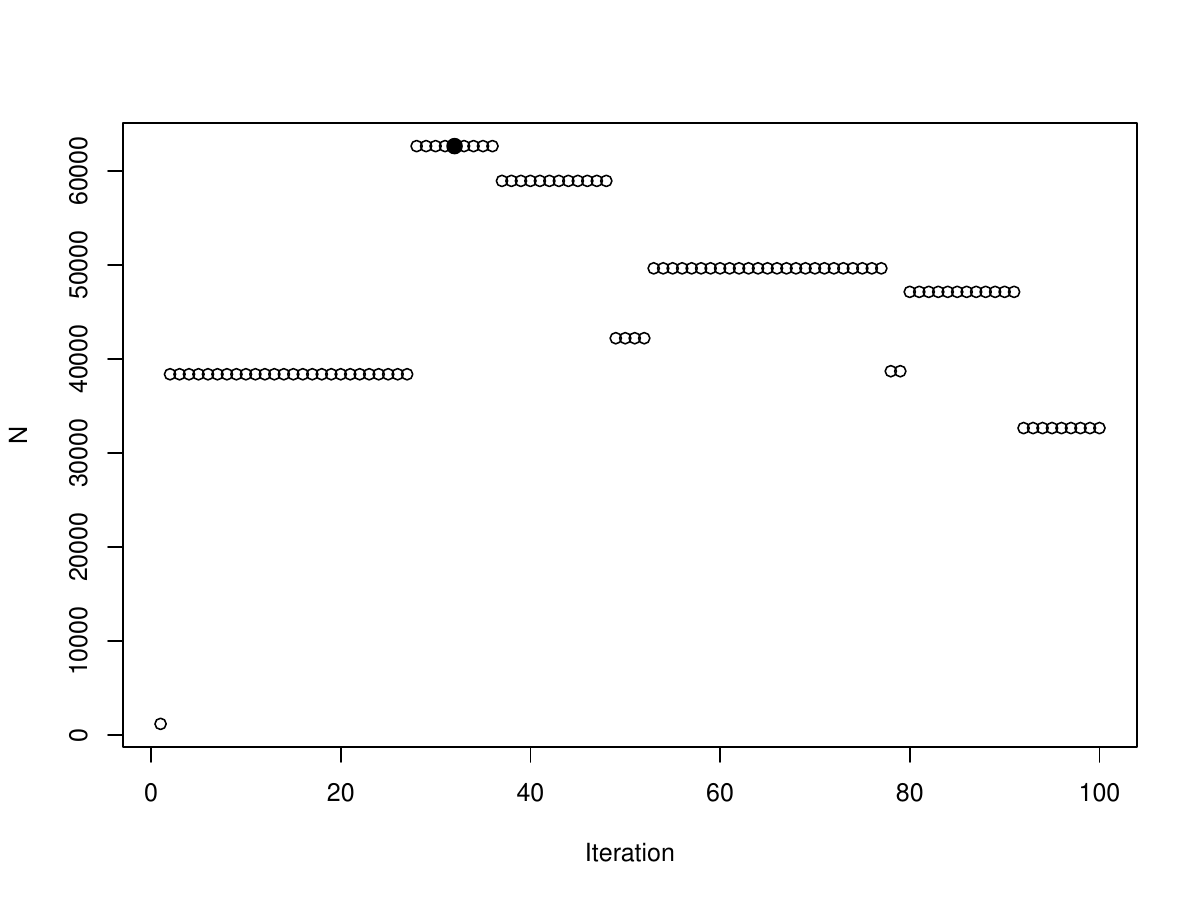}
  \end{minipage}
    \caption{{  Illustration of the algorithm for obtaining the identification region of the population size $N$ for the Engage Montreal dataset. On the left (on the right), the iterations that led the lower (upper) bound of $N$, where the black circle represents the global minimum (maximum). The corresponding identification region is [3969; 62658].}}
\label{fig:boundsEngage}
\end{figure}

Participants who received coupons for peer recruitment subsequently reported the number $C^*_i$ that they successfully distributed, $i=1,\dots, 1179$. Comparing the self-reported number of distributed coupons to the number of recruits linked to each recruiter can help to empirically investigate potential violations of Assumptions \ref{assum:givecoup1} and \ref{assum:givecoup2}. Descriptive statistics show that 47 participants recruited fewer GBM members than the self-reported number of coupons that were distributed; this is a practical violation of Assumption \ref{assum:givecoup2}. Also, 41 participants distributed fewer coupons than the minimum number of unrecruited neighbours who were available for recruitment, which is a clear violation of Assumption \ref{assum:givecoup1}. This indicates that coupon distribution and return policies are most likely violated for the Engage dataset, and that point estimation and inference for the number of GBM members may not be valid.
%

We implement a stochastic optimization algorithm to derive the identification region for the Montreal GBM population. To find the global minimum of $N$, we set $\epsilon=0.2$ and started with the proposal $\bm{\tilde{y}}=(\tilde{d}_1,\dots, \tilde{d}_{1179})$; for the global maximum, we set $\epsilon=1.2$ and use the starting proposal $\bm{\tilde{y}}=(d_1-1,\dots, d_{1179}-1)$. Across $t=1, 2, \dots, 100$ iterations, the algorithm yielded the region [3969; 62658] as the smallest that contains the true GBM population living in the Metropolitan area of Montreal. This is illustrated in Figure \ref{fig:boundsEngage}. Applying our inferential framework to the observed Engage data yields $\hat{N}=6683$, with a 95\% confidence interval of $[6340; 7044]$.

\section{Discussion}
We have proposed a frequentist, model-based approach for estimating the size of a hidden population using a single RDS survey. In doing so, we assumed that the probability that a new recruit (turned recruiter) successfully brings in a given number of participants depends on both the number of unrecruited neighbours at the time of recruitment and the probability that neighbours who accept coupons enter the study. This yields the insight that the number of coupons that a recruiter successfully distributes may reveal partial or complete information regarding the number of unrecruited neighbours at the time of recruitment. In particular, a recruiter who receives $C$ coupons but successfully distributes $C^*<C$ has exactly $C^*$ neighbours available for recruitment; a recruiter who distributes $C^*=C$ coupons may have more than $C$ unrecruited neighbours at the time of recruitment. This yields observations that are either random realizations from the target probability distribution or structurally censored on the right. Under a network-based approach, we proposed an inferential framework for the population size $N$; we characterized the asymptotic behavior of our proposed population size estimator and assessed its small sample performance under departures from model assumptions.

We applied our methodology to an RDS dataset from a cohort study about the overall sexual health and behaviors of the GBM community that took place in the cities of Montréal, Toronto, and Vancouver. In Montréal, the estimate of 6683 is far smaller than the expected number of GBM members residing in the Metroplotian area according to recent reports, implying possible violations of the coupon distribution assumptions.  In fact, research conducted by the Public Health Agency of Canada \citep{sorge2023estimation} and Institut de la statistique du Québec \citep{INSPQ2021} provided estimates of 3.2\% and 4.2\% respectively for the proportion of GBM aged 15 years or older in Quebec; given that approximately half of the population of the province resides in the Metropolitan area of Montreal, the corresponding population size estimates are 57781 and 75838 respectively. Note that there is a slight limitation of comparability with our estimates due to age cutoff because the Engage study recruited GBM members aged 18 years or older.

Violations of the coupon distribution assumptions imply that the observed data are not random realizations from the target probability distribution; an unknown proportion of the observations are measured with error. This missing data problem renders the proposed inferential framework for $N$ invalid as one can construct two hypothetical full data versions that give rise to the same observed data distribution given the vectors of self-reported degrees and recruitment times. We developed a methodology for deriving the identification region of $N$, i.e., the smallest interval that contains $N$, by constructing a stochastic optimization algorithm that explores the set of hypothetical full data versions of the observed data to obtain the global minimum and maximum of $N$. The optimization technique was evaluated using simulated RDS data, then applied to a real-word RDS dataset to obtain the lower and upper bounds for the number of GBM members in the Metropolitan area of Montréal.

Crucial limitations for the proposed inferential framework include the importance for assessing violations of coupon distribution and acceptance assumptions. As highlighted in the sensitivity analysis, severe violations of model assumptions may induce substantial bias in the population size estimate as well as undercoverage of the 95\% confidence interval for $N$. Further, it may not be possible to check violations of the coupon distribution and acceptance assumptions since participants do not usually report the number of coupons that they successfully distributed. Future works using RDS methods may consider soliciting this information from recruiters. However, when such information is unavailable, we recommend the use of existing data regarding the structure of the target population to validate the population size estimate and the associated confidence interval given the observed data. If the point estimate is not consistent with existing data/knowledge about the population, we recommend the implementation of the optimization algorithm to derive bounds for $N$.

\bigskip
\begin{center}
{\large\bf SUPPLEMENTARY MATERIAL}
\end{center}

\begin{description}

\item[Proofs of consistency and asymptotic normality for the MLE of $N$:] This document presents proofs of all Propositions and Theorems for when $\rho$ is unknown, as well as additional details regarding the optimization algorithm. (.pdf file)

%

\end{description}


\end{document}